\documentclass[pra,amsmath,letterpaper,floatfix]{revtex4}

\usepackage{amsmath}
\usepackage{amssymb}
\usepackage{graphicx}
\usepackage{color}

\newcommand{\Ntwo}{N$_{2}$ }

\begin{document}

\title{Control of molecular rotation in the limit of extreme rotational excitation}

\author{V.~Milner and J.~W.~Hepburn}
\date{\today}

\affiliation{Department of  Physics \& Astronomy, The University of British Columbia, Vancouver, Canada}

\begin{abstract}
Laser control of molecular rotation is an area of active research. A number of recent studies has aimed at expanding the reach of rotational control to extreme, previously inaccessible rotational states, as well as controlling the directionality of molecular rotation. Dense ensembles of molecules undergoing ultrafast uni-directional rotation, known as molecular superrotors, are anticipated to exhibit unique properties, from spatially anisotropic diffusion and vortex formation to the creation of powerful acoustic waves and tuneable THz radiation. Here we describe our recent progress in controlling molecular rotation in the regime of high rotational excitation. We review two experimental techniques of producing uni-directional rotational wave packets with a ``chiral train'' of femtosecond pulses and an ``optical centrifuge''. Three complementary detection methods, enabling the direct observation, characterization and control of the superrotor states, are outlined: the one based on coherent Raman scattering, and two other methods employing both resonant and non-resonant multi-photon ionization. The capabilities of the described excitation and detection techniques are demonstrated with a few examples. The paper is concluded with an outlook for future developments.
\end{abstract}

\maketitle

\section{Introduction}
The study of the properties of highly excited molecules has yielded great insight into the details of chemical dynamics. While much of the work has focused on energy partitioning in the products of photodissociation or reactions \cite{Eppink1999, Polanyi1987}, a great deal of insight has also been obtained through analyzing the vector correlations in the reaction products (typically rotation-translation correlations) \cite{Hall1989, Burak1987}.  There has also been a strong focus on the initial states of molecules, particularly initial vibrational states. An important element of this work has been to use initial state preparation to control chemical reactions \cite{Zare1998, Crim1996},  but there has also been significant interest in the relaxation of these excited molecules, either unimolecular or collisional \cite{Crim1996, Silva2001, Flynn1996}, and in the spectroscopy of highly excited states near isomerization thresholds \cite{Ishikawa1999}.

Most of the work to date in initial state preparation has involved electronic excitation to metastable excited states, where state preparation of specific ro-vibronic states is possible \cite{Burak1987, Moore1983}, or excitation of vibrationally excited levels of the ground electronic state by stimulated emission pumping through using an excited electronic state \cite{Silva2001} or stimulated Raman adiabatic passage \cite{Vitanov2001}. All these techniques are very efficient because they take advantage of relatively low order transitions with strong transition probabilities. The other advantage of electronic or vibrational state preparation is that it deposits a significant amount of internal excitation in the molecule, with a resulting strong impact on the subsequent dynamics.

In contrast to final product state distributions, where a great deal has been learned from studying rotational states and vector correlations \cite{Polanyi1987, Hall1989, Burak1987}, relatively little work has been done using initial rotational state preparation. The main reason for this is the major difficulty of producing significant amounts of rotational excitation, where selection rules limit the ability to excite large changes in the rotational quantum number through optical excitation. One method that has been used to study the relaxation of highly excited rotational levels is to produce broad distributions of rotationally excited products from reactions \cite{Polanyi1972} or photodissociation \cite{Ho1982}. Specific rotational levels of moderate excitation can also be prepared through excitation of hot samples to fluorescing excited states \cite{Saenger1983}.

Strong electric fields of ultrashort laser pulses offer a new approach to preparing molecules in well defined and highly excited rotational levels, either through impulsive excitation with a series of pulses, or through adiabatic excitation using frequency chirping. Both of these techniques are described in this chapter, using some examples from our recent work. Because these methods of excitation result in a rotational wave packet with well defined and variable properties, it is possible to create not only very highly excited rotational states with a few eV of rotational excitation, but the range of excited rotational levels can be also rather narrow.

These highly rotationally excited molecules have many interesting properties, which opens up a number of possible avenues for study. In addition to the well defined rotational energy, the extreme rotational excitation distorts the molecular frame which can have an impact on dynamics and spectroscopy, and the large rotational spacing for these highly excited states makes them resistant to collisional relaxation or reorientation.

\section{The challenge of high rotational excitation}
Strong non-resonant laser fields affect molecular rotation by exerting an angle dependent torque on the field induced molecular dipole \cite{Zon75, Friedrich95, Seideman95, Rosca02, Dooley03, Underwood05}. The interaction is described by the potential
\begin{equation}\label{eq-potential}
    U(\theta, t)=-\frac{1}{4}\Delta \alpha E^2(t)\cos^2{\theta},
\end{equation}
where $\Delta\alpha$ is the anisotropy of the molecular polarizability, $E(t)$ is the electric field envelope, and $\theta$ is the angle between the molecular axis and field polarization. In the limit of ultrashort laser pulses, the pulse acts on a molecule as an instantaneous rotational ``kick''. In the quantum picture, a laser kick induces multiple Raman transitions between the rotational states of the molecule, transferring population from lower to higher rotational states. How high a molecule ``climbs'' up the rotational ladder is determined by the typical amount of angular momentum (in units of $\hbar$) transferred from the field to the molecule, the so-called ``kick strength'' \cite{Averbukh2001}, defined as
\begin{equation}\label{eq-kick}
    P=\frac{\Delta\alpha}{4\hbar}\int E^2(t)dt.
\end{equation}

\begin{figure}[h]
\includegraphics[width=1\columnwidth]{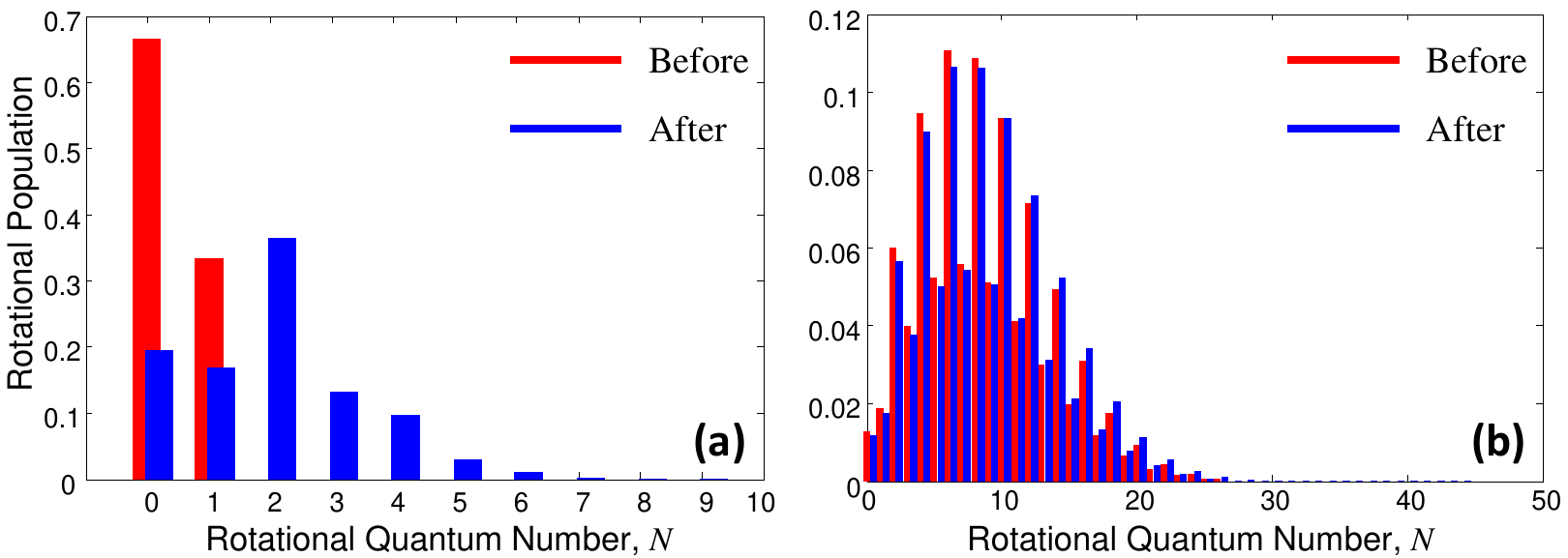}
\caption{(Color online) Numerically calculated distribution of rotational population in $^{14}$N$_{2}$ before and after the interaction with a single linearly polarized femtosecond pulse of duration 120 fs and peak intensity $10^{13}$ W/cm$^{2}$. (\textbf{a}) Zero temperature. (\textbf{b}) Room temperature.}
\label{fig-kick}
\end{figure}

Despite the success of rotational excitation with a single non-resonant femtosecond laser pulse, the approach suffers from a limited capacity of reaching high rotational states. To avoid molecular ionization and plasma breakdown in dense gas samples, the peak intensity of the excitation pulse must not exceed $\approx 10^{13}$ W/cm$^{2}$. Even for a molecule with a relatively high polarizability anisotropy, such as \Ntwo with $\Delta \alpha = 0.7$ {\AA}$^{3}$, this intensity results in a kick strength $P=3.6$ and correspondingly small change in the distribution of rotational population. This is illustrated in Fig. \ref{fig-kick}, where the rotational distribution of nitrogen molecules (both at 0K and room temperature) is plotted before and after the interaction with a single 120~fs, $10^{13}$~W/cm$^{2}$ pulse. In room temperature ensembles, the effect of the single pulse on the rotational distribution is hardly visible.

\section{Non-adiabatic rotational control with trains of ultrashort laser pulses}

\subsection{Selective rotational excitation via quantum resonance.}
\begin{figure}[b]
\includegraphics[width=1\columnwidth]{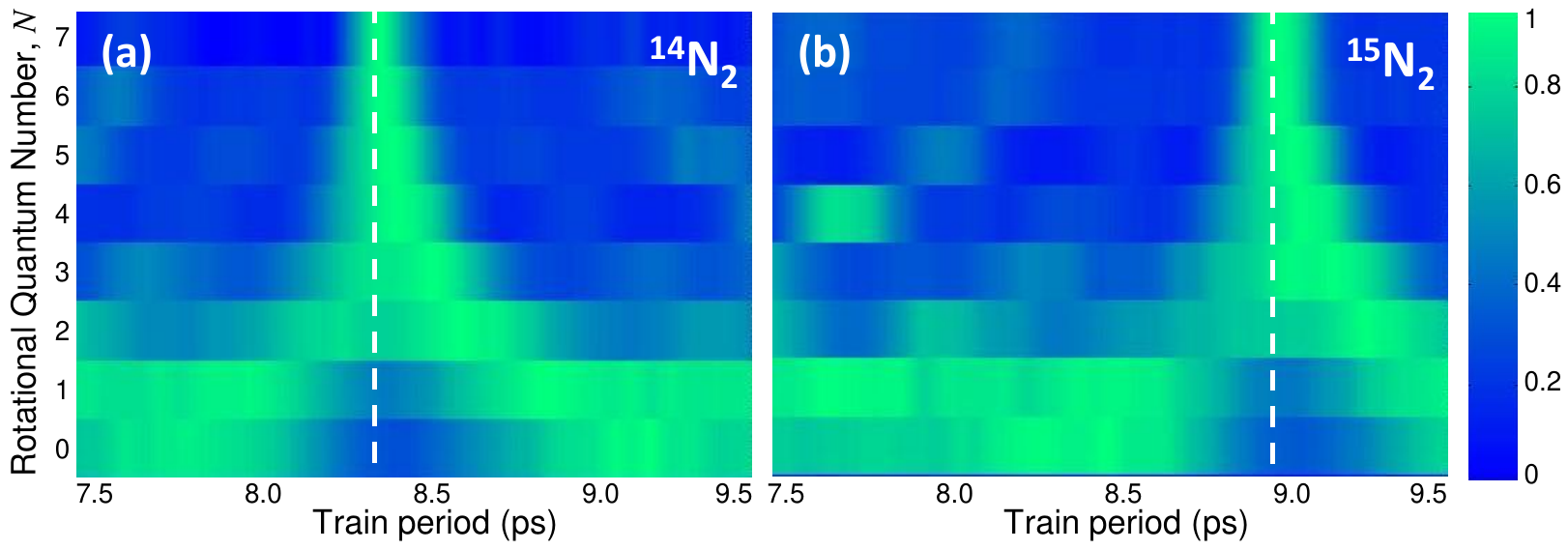}
\caption{(Color online) Experimentally measured rotational population (color coded) of the first eight rotational levels in $^{14}$N$_{2}$ (\textbf{a}) and $^{15}$N$_{2}$ (\textbf{b}). For a pulse train period equal to the revival time (white vertical lines) the population is efficiently transferred from the initial states $N=0,1,2$ to higher states $N=3, 4,...7$.}
\label{fig-train}
\end{figure}

Splitting a single strong laser pulse into a series of weaker pulses, a ``pulse train'', helps mitigate the problem of strong-field ionization \cite{Leibscher03, Cryan09}. If the pulses in the train are separated by an integer multiple of the molecular revival time ($T_\text{rev}=1/(2Bc)$, where $B$ is the rotational constant of a molecule in units of cm$^{-1}$ and $c$ is the speed of light in units of cm/s), the effect of the train is identical to that of a single pulse of kick strength $P_{\text{train}}=\sum_n P_{n}$, with $P_{n}$ being the kick strength of the $n$-th individual pulse in the train. For a sequence of laser pulses of equal intensity, the molecular angular momentum grows linearly with the pulse number. This condition of the so-called \textit{quantum resonance} has been recently realized and studied experimentally \cite{Zhdanovich12}. The results, shown in Fig.\ref{fig-train}, confirmed the enhancement of rotational excitation at quantum resonance and demonstrated its utility for selective rotational excitation in molecular mixtures.

The method of exciting molecular rotation with femtosecond pulse trains, though very powerful in terms of its selectivity, runs into two obstacles on the way to reaching high rotational states. The first one is of a technical nature. Producing sequences of multiple laser pulses becomes exceedingly difficult with increasing time separation between the pulses. A train of eight equal-amplitude pulses separated by the revival time of a nitrogen molecule (8.4 ps) requires either a complicated setup of multiple nested interferometers \cite{Cryan09} or an amplitude pulse shaper. The energy throughput of the latter drops drastically with the increasing number of pulses, thus limiting the cumulative kick strength of the train to rather low values. The second impediment stems from the centrifugal distortion which modifies the revival time and effectively de-tunes the molecule from the quantum resonance, thus suppressing the efficiency of the rotational excitation \cite{Floss2012}.

\begin{figure}[h]
\includegraphics[width=1\columnwidth]{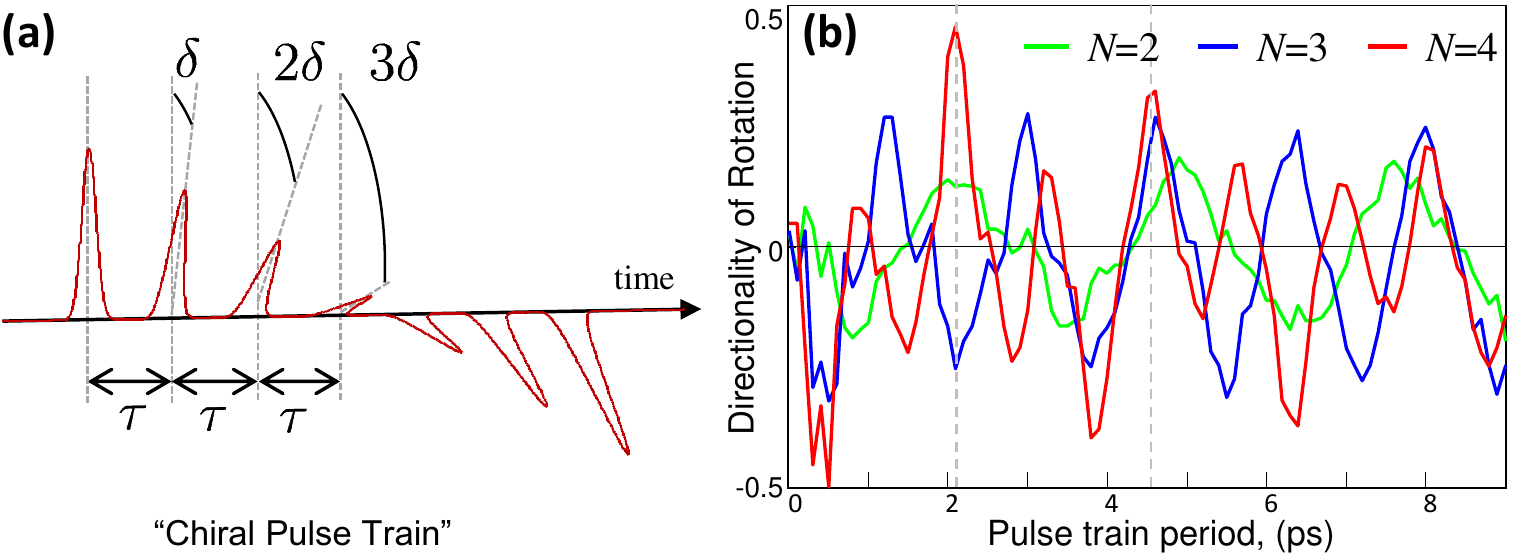}
\caption{(Color online) (\textbf{a}) ``Chiral train'' of femtosecond pulses, separated by a constant time period $\tau $. The vector of linear polarization rotates from pulse to pulse by a constant angle $\delta $. (\textbf{b}) Experimentally measured rotational directionality of nitrogen molecules occupying rotational levels $N=2,3$ and 4 (green, blue and red curves, respectively). Multiple rotational states ($N\leq7$) are excited by the same pulse train, whose time period dictates whether rotation of different states, e.g. $N=3$ and $N=4$, proceeds in the same (right dashed line) or opposite (left dashed line) direction.}
\label{fig-chiral}
\end{figure}

Femtosecond pulse trains offer another control knob to rotational excitation of molecules, namely the control of its directionality. By applying a series of laser pulses, linearly polarized at an angle with respect to one another, either clockwise or counter-clockwise rotation can be initiated in a molecular ensemble\cite{Fleischer09}. The effect has been experimentally demonstrated with a sequence of only two pulses \cite{Kitano09}, and longer pulse trains \cite{Zhdanovich11, Bloomquist12}. The latter were dubbed ``chiral'' because of the directional rotation of light polarization from pulse to pulse shown in Fig.\ref{fig-chiral}(\textbf{a}). Chiral pulse trains enable control of the sense of molecular rotation with state selectivity. As illustrated in Fig.\ref{fig-chiral}(\textbf{b}), the train period can be set so as to result in the same (right dashed line) or opposite (left dashed line) rotational directionality for molecules in $N=3$ and $N=4$ rotational states.

\section{Adiabatic rotational control with an optical centrifuge}

\subsection{Optical centrifuge.}
\begin{figure}[b]
\includegraphics[width=1\columnwidth]{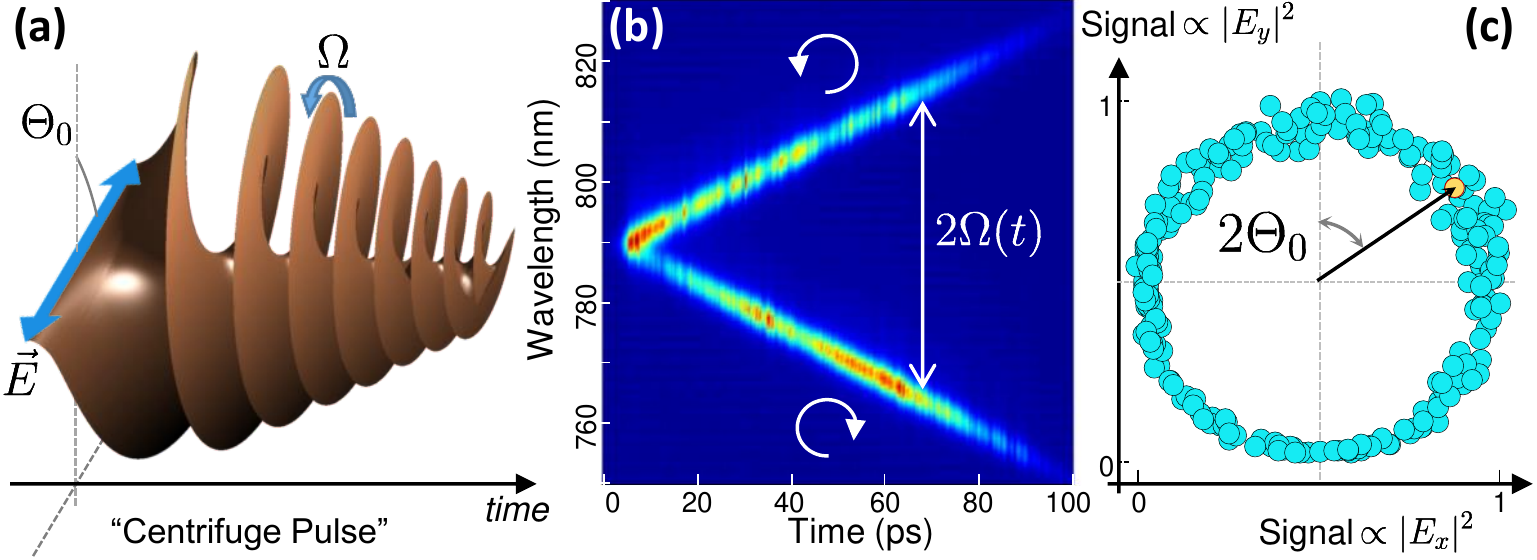}
\caption{(Color online) (\textbf{a}) Illustration of a ``centrifuge pulse''. The vector of linearly polarized laser field $\vec{E}$ undergoes rotation with an angular frequency $\Omega $, linearly growing in time (number of rotations is grossly reduced for clarity). $\Theta_{0}$ is the orientation angle of the centrifuge. (\textbf{b}) Experimentally measured time-frequency spectrogram of the centrifuge pulse. The two spectral components are circularly polarized (opposite sense) and frequency chirped (opposite sign). (\textbf{c}) Correlation plot of the two orthogonal projections of the centrifuge field $\vec{E}$, reflecting the random distribution of the orientation angles from pulse to pulse.}
\label{fig-centrifuge}
\end{figure}

A number of alternative approaches to controlling molecular rotation has been suggested \cite{Karczmarek99,Li00,Vitanov04,Cryan11}. An efficient method of accelerating molecular rotation with an ``optical centrifuge'' \cite{Karczmarek99} has been successfully implemented \cite{Villeneuve00,Yuan11,Korobenko14a}. Molecular spinning with an optical centrifuge is achieved by forcing the molecules to follow the rotating potential of Eq.\ref{eq-potential} created by the rotating polarization of a laser field $\vec{E}$, as schematically depicted in Fig.\ref{fig-centrifuge}(\textbf{a}).

The field of an optical centrifuge is produced from regeneratively amplified ultrashort laser pulses of 35 fs duration. Frequency chirps of equal magnitude and opposite sign are applied to the two spectral halves of the original pulse by means of standard double-grating pulse stretchers \cite{Milner14b,Korobenko14c}. The two spectral components are then polarized with an opposite sense of circular polarization and re-combined in space and time. Experimentally measured time-frequency spectrogram of the centrifuge field is shown in Fig.\ref{fig-centrifuge}(\textbf{b}). Interference of the two centrifuge arms (upper and lower traces on the spectrogram) results in a linearly polarized field, whose plane of polarization undergoes an accelerated rotation. The frequency of rotation $\Omega$ is determined by the instantaneous frequency difference, $2\Omega(t) $, between the two spectral components, and can reach 10 THz by the end of a 100 ps long centrifuge pulse.

Due to the lack of interferometric stability between the two long arms of the centrifuge pulse shaper, the initial orientation angle of the centrifuge polarization with respect to the laboratory frame, denoted $\Theta _{0}$ in Fig.\ref{fig-centrifuge}(\textbf{a}), changes randomly from pulse to pulse. To fully characterize the centrifuge, $\Theta_0$ can be found by measuring the two orthogonal projections ($E_{x}$ and $E_{y}$) of the field vector for every centrifuge pulse by means of the nonlinear optical gating technique \cite{Korobenko14c}. The angle is extracted from the correlation plot of two signals proportional to $|E_{x}|^2$ and $|E_{y}|^2$, as shown in Fig.\ref{fig-centrifuge}(\textbf{c}).

\subsection{Direct detection of molecular superrotation.}
Controlling molecular rotation under ambient conditions, i.e. room temperature and atmospheric pressure, offers an appealing pathway towards laser control of chemical reactions. Yet the very high laser intensity required for an angular confinement of room-temperature molecules in a dipole potential of Eq.\ref{eq-potential} (e.g. $2.7\times 10^{13}$ W/cm$^{2}$ in the case of nitrogen) brings about a number of strong-field effects, such as multi-photon ionization and plasma breakdown. Their influence on the molecular ensemble must be carefully distinguished from the effects of true rotational excitation. To that end, three direct methods of detecting and analyzing the rotation of centrifuged molecules have been developed and are discussed in detail below.
\begin{figure}[b]
\includegraphics[width=1\columnwidth]{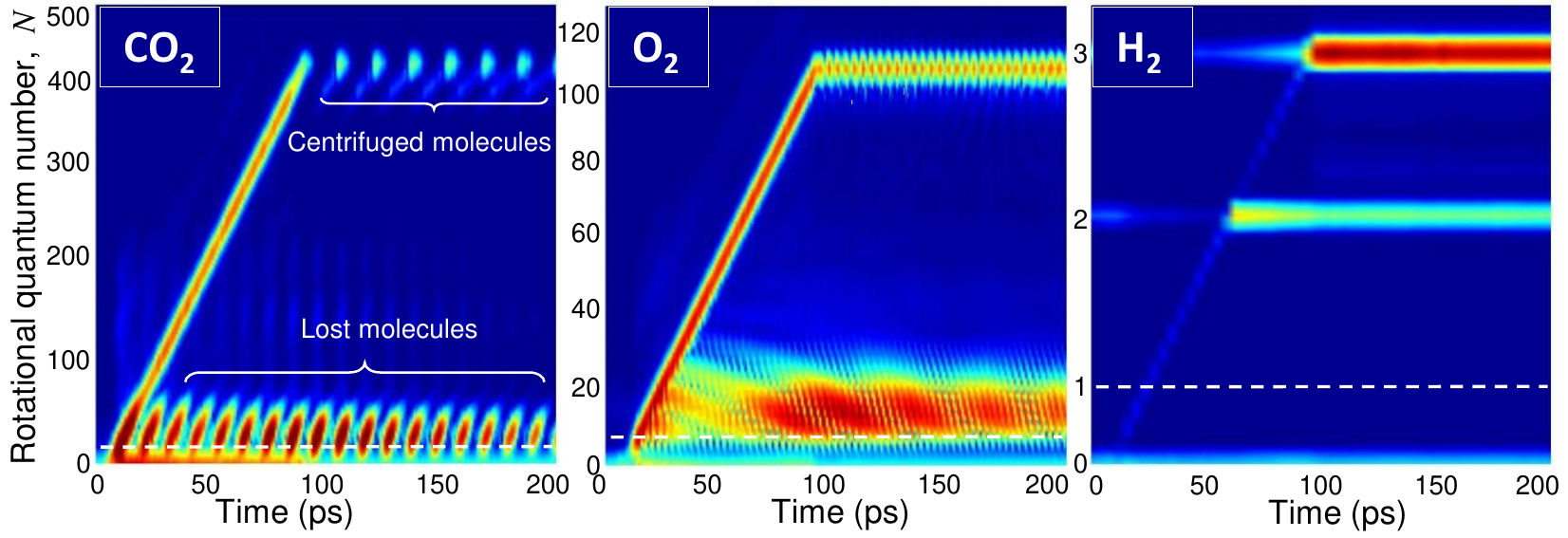}
\caption{(Color online) Raman spectra of centrifuged CO$_{2}$, O$_{2}$ and H$_{2}$, as a function of the probe delay with respect to the beginning of the centrifuge pulse. Tilted traces correspond to the linearly increasing Raman shift due to the accelerated rotation of molecules inside the centrifuge. In the case of CO$_{2}$ and O$_{2}$, upper and lower oscillatory traces stem from the coherent dynamics of the centrifuged molecules and molecules lost from the centrifuge, respectively. Horizontal dashed lines indicate the most populated rotational level at room temperature. Color coding is used to reflect the signal strength in logarithmic scale.}
\label{fig-raman}
\end{figure}

\subsubsection{Raman scattering from the ensemble of centrifuged molecules.}
Synchronous molecular rotation corresponds to a coherent superposition of a few rotational quantum states - a ``rotational wave packet'', with an average frequency separation matching the frequency of the classical rotation. Owing to the time-dependent wave packet coherence between the quantum states separated by $\Delta N=\pm2$ (with $N$ being the rotational quantum number), the spectrum of a probe laser beam passing through the ensemble of centrifuged molecules develops Raman sidebands. The magnitude of the Raman shift equals twice the rotation frequency, while its sign reflects the direction of molecular rotation with respect to the probe's circular polarization \cite{Korobenko14a}. In classical terms, the frequency shift can also be viewed as a result of the rotational Doppler effect \cite{Korech13}.

Fig.\ref{fig-raman} shows two-dimensional Raman spectrograms of centrifuged CO$_{2}$, O$_{2}$ and H$_{2}$ molecules, where the Raman spectrum of the scattered probe light is plotted as a function of the probe delay with respect to the beginning of the centrifuge pulse. The vertical scale has been converted from the measured energy shift $\Delta E$ to the rotational quantum number $N$ according to $\Delta E=B N (N+1) - D N^2 (N+1)^2$, with $B$ and $D$ being the rotational and centrifugal constants of a molecule, respectively. During the first 90 ps, CO$_{2}$ and O$_{2}$ molecules are trapped in the centrifuge potential and faithfully follow its accelerated rotation. Free evolution of the created rotational wave packet starts after the molecules are released from the centrifuge and appears as a series of periodic revivals, indicative of the coherent rotational dynamics. The revival time of a heavier carbon dioxide (excited to $N\gtrsim 400$) is much longer than that of a lighter oxygen molecule (occupying rotational states around $N=100$). Coherent Raman signal at lower $N$ values corresponds to the molecules which were too hot to follow the centrifuge and spilled out of it at the beginning of the spinning process. The rotational levels of even lighter hydrogen are so far apart that the centrifuge is capable of driving only two rotational transitions, $N=0 \rightarrow N=2$ in parahydrogen and $N=1 \rightarrow N=3$ in orthohydrogen (two horizontal traces in the right panel of Fig.\ref{fig-raman}).
\begin{figure}[b]
\includegraphics[width=1\columnwidth]{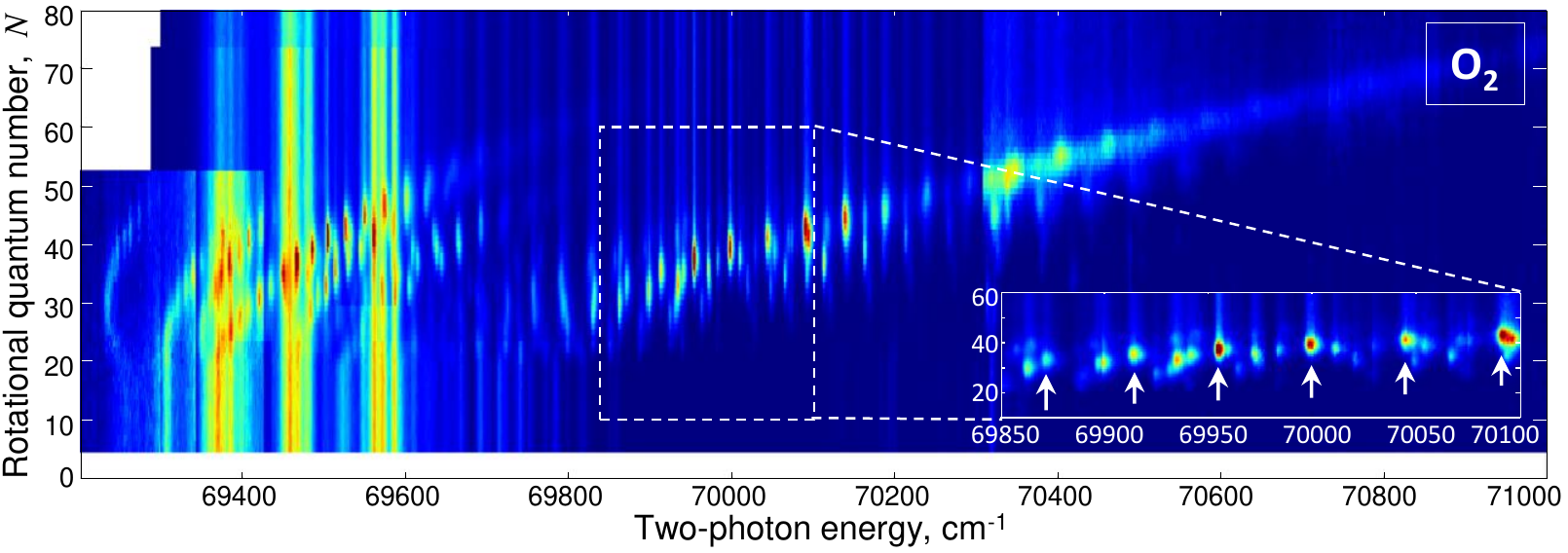}
\caption{(Color online) 2D REMPI spectrogram of the centrifuged O$_{2}$ molecules. Ion signal is color coded as a function of the frequency of the ionizing laser field, which undergoes two-photon absorption on the  $C^3\Pi_g(v'=2)\leftarrow\leftarrow X^3\Sigma_g^-(v''=0)$ transition, and molecular angular momentum, which is defined by the terminal angular frequency of the centrifuge. White arrows in the inset point at sharp REMPI resonances corresponding to $N=31,33,35,37,39$ and 41.}
\label{fig-rempi}
\end{figure}

\subsubsection{Resonant photo-ionization spectroscopy.} \label{sec-rempi}
Although very efficient in detecting rotational coherence and evaluating the degree of rotational excitation, Raman spectroscopy cannot provide the direct measure of rotational population and lacks the sensitivity for detecting very low density samples. Resonance-enhanced multi-photon ionization (REMPI) is often employed for this purpose due to its high sensitivity, spectral resolution and versatility. However, the technique relies on good understanding of the rotational spectrum of a molecule in the excited electronic state. In the case of extreme rotational excitation by an optical centrifuge, the assignment of multi-photon resonances and their strengths can be particularly challenging. Yet the ability to control the rotational frequency of a molecule makes it possible to identify the spectroscopic lines, and hence to determine the population distribution among the rotational states \cite{Korobenko14b}. Fig.\ref{fig-rempi} shows an example of applying the technique of ``centrifuge spectroscopy'' to oxygen superrotors. Adding the second dimension - a well defined rotational quantum number $N$, to the REMPI scan enables one to separate and interpret multiple rotational branches, visible on the plot as a set of parabolic traces.
\begin{figure}[b]
\includegraphics[width=1\columnwidth]{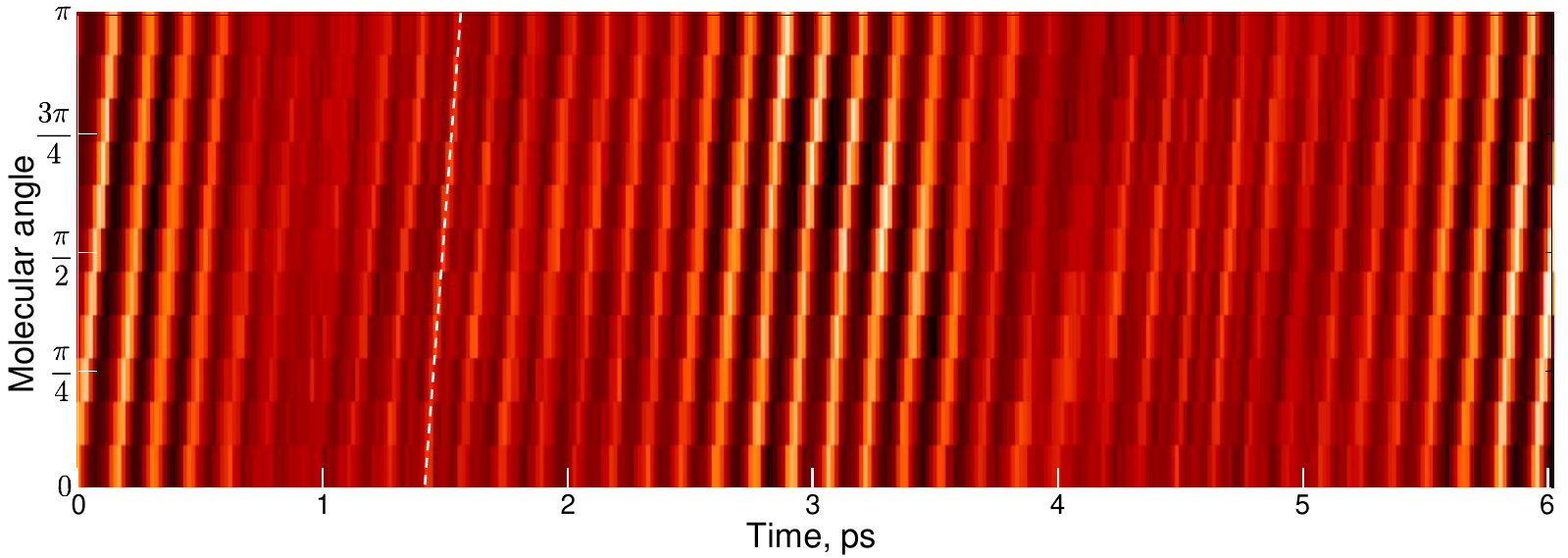}
\caption{(Color online) Probability density of the centrifuged oxygen gas as a function of the molecular angle and the free propagation time, i.e. the time elapsed since the molecules have been released from the centrifuge. The white dashed line (around 1.5 ps) marks the calculated trajectory of a ``dumbbell'' distribution rotating with the classical rotational frequency of an oxygen molecule with an angular momentum of 39$\hbar$.}
\label{fig-cogwheel}
\end{figure}

\subsubsection{Wave packet imaging.}
Yet another direct way of detecting molecular superrotation is based on the method of velocity map imaging (VMI) \cite{Larsen98}. In VMI, an intense femtosecond laser pulse dissociates the molecule via a multi-photon ionization process followed by a Coulomb explosion. The recoiling ions are extracted with an electric field and projected onto a micro-channel plate detector equipped with a phosphorus screen, where they leave spatially resolved fluorescence traces. The method enables one to map out the direction of the ion recoil and, therefore, the angular distribution of molecular axes at the moment of explosion \cite{Dooley03}. In Fig.\ref{fig-cogwheel}, the orientation angle of oxygen molecules, which have been spun up to $N=39$ in an optical centrifuge, is plotted as a function of the delay between the time of release from the centrifuge and the time of Coulomb explosion \cite{Korobenko14c}. Classical-like rotation of the molecular wave function is reflected by the series of equally tilted lines, whose slope is equal to the classical rotational frequency of an oxygen molecule with an angular momentum of 39$\hbar$. Since the created rotational wave packet consists of more than two rotational eigenstates, it undergoes an angular dispersion which results in the periodic smearing of the VMI signal.

\subsection{Control of centrifuge-induced rotation.}
\begin{figure}[h]
\includegraphics[width=1\columnwidth]{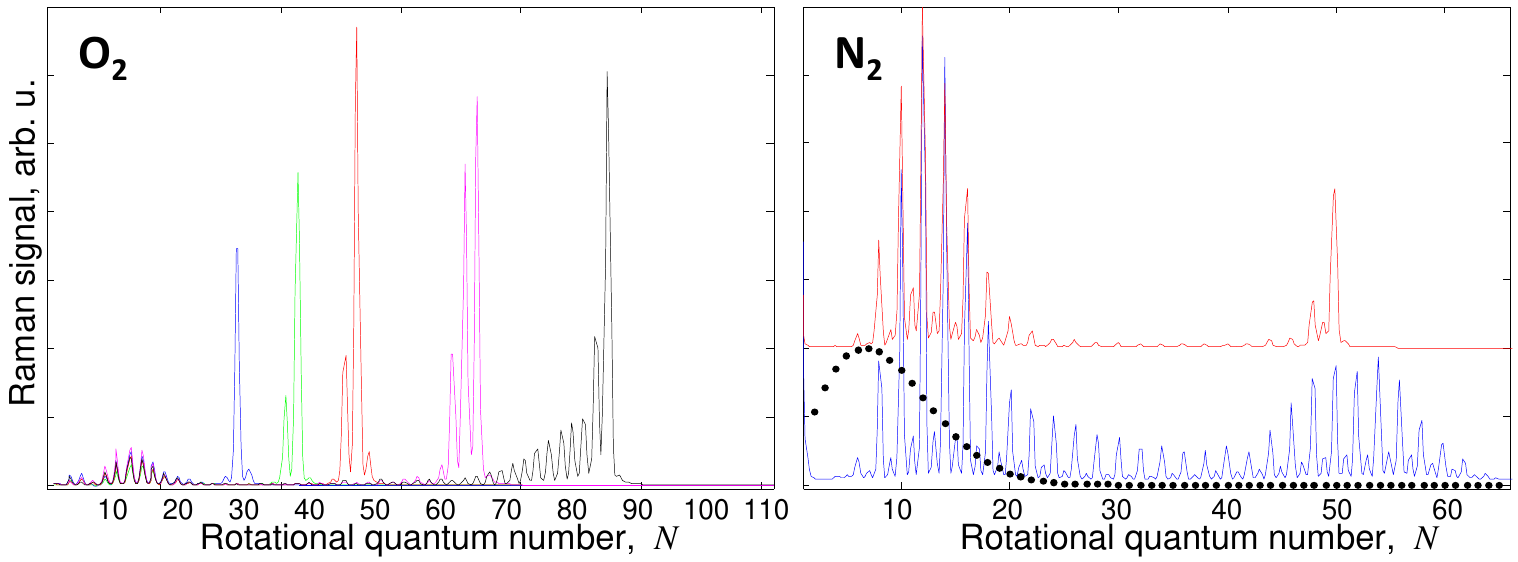}
\caption{(Color online) Examples of controlling rotational wave packets with an optical centrifuge. Shifting the center of a narrow wave packet is shown for centrifuged oxygen on the left, whereas varying the width of the wave packet is demonstrated for nitrogen on the right. Black dots represent the thermal distribution at room temperature.}
\label{fig-control}
\end{figure}

The great value of an optical centrifuge stems not only from its capacity to access extreme rotational states, but also from an equally important controllability over the created rotational wave packets. Described earlier in Section \ref{sec-rempi}, this property has been utilized in the new approach to rotational spectroscopy, where the ability to excite molecules to a specific well-defined rotational state, rather than a broad distribution of states, greatly simplifies the interpretation of the detected REMPI resonances. In Fig.\ref{fig-control}, this power of controlled centrifuge excitation is demonstrated with frequency-resolved coherent Raman scattering. Unlike the time-resolved Raman spectra shown earlier in Fig.\ref{fig-raman}, here the detection was executed with spectrally narrow (3.75 cm$^{-1}$, FWHM) probe pulses. As a result, individual rotational states are resolved (at least in light O$_{2}$ and N$_{2}$ molecules) and the measured spectrum reveals the exact composition of the created rotational wave packet. On the left, the controlled shift of a narrow wave packet, consisting of only a few rotational eigenstates, across an extremely broad range of angular momenta from $N=0$ to $N\approx 100$ is demonstrated in centrifuged oxygen. On the other hand, it is often useful to simultaneously populate and monitor as many rotational states as possible, for instance in the study of rotational de-coherence in dense media \cite{Milner14a}. The centrifuge technique can achieve this goal too, as shown in the right panel of Fig.\ref{fig-control} which depicts the transition from a narrow to ultra-broad rotational wave packet created in nitrogen molecules. The latter is composed of more than 60 rotational states.

\subsection{Dynamics of molecular superrotors.}
\subsubsection{Effects of centrifugal distortion on rotational dynamics.}
\begin{figure}[b]
\includegraphics[width=1\columnwidth]{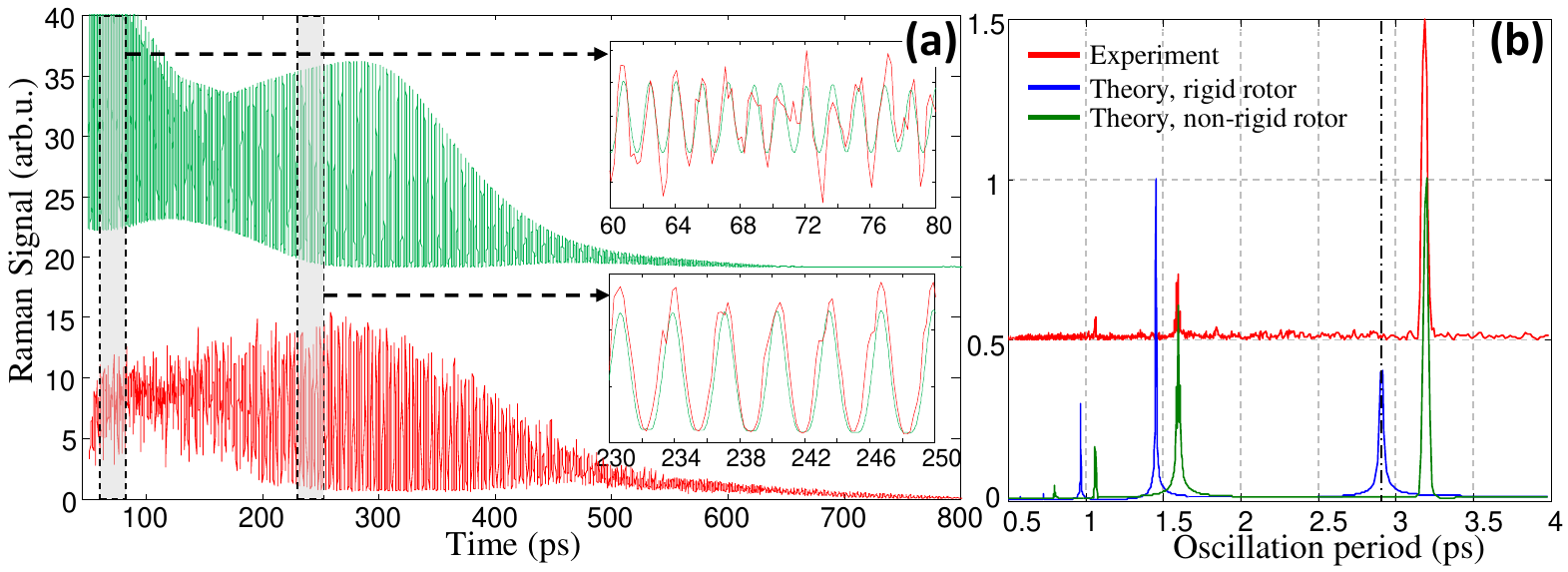}
\caption{(Color online) (\textbf{a}) Experimentally detected (lower red) and numerically calculated (upper green) evolution of a superrotor wave packet centered at $N=69$ in oxygen. Two insets illustrate the agreement between the two curves at different time moments. (\textbf{b}) Fourier transform of the collected data (upper red) and the numerically simulated wave packet oscillations without and with the centrifugal distortion taken into account (blue and green curves, respectively).}
\label{fig-dynamics}
\end{figure}

As illustrated earlier in the text (Fig. \ref{fig-raman}), molecules released from the centrifuge generate an oscillatory Raman signal, characteristic of the coherent rotation with well defined relative phase relation between the quantum states inside a rotational wave packet. Time-resolved coherent Raman response from a wave packet centered at $N=69$ in oxygen is plotted at the bottom of Fig.\ref{fig-dynamics}(\textbf{a}). Knowing the wave packet composition from the state-resolved detection discussed above, one can model the rotational dynamics numerically (upper green curve). Owing to the quadratic scaling of energy with the rotational quantum number $N$, the oscillations exhibit multiple commensurate time periods. In the example used here, the initial period of 1.6 ps changes to twice that value at later times, in agreement with the numerical calculations [two insets in Fig.\ref{fig-dynamics}(\textbf{a})].

As follows from the general theory of quantum wave packets \cite{Leichtle96,Seideman99}, the oscillation periods are inversely proportional to the second derivative of $E(N)$ with respect to $N$. For a rigid-rotor model of oxygen, this results in the main period equal to the quarter of the rotational revival time, i.e. $T=(8Bc)^{-1}\approx 2.9$ ps, marked by the vertical dash-dotted line in Fig.\ref{fig-dynamics}(\textbf{b}). Centrifugal distortion pulls the atoms apart, lengthening the molecular bond and increasing the molecular moment of inertia \textbf{I}. Being inversely proportional to \textbf{I}, the rotational constant $B$ decreases, resulting in a longer period of oscillations at higher levels of rotational excitation, $T=\left[8Bc(1-6\epsilon N(N+1))\right]^{-1}$, where $\epsilon \equiv D/B\approx 3\times 10^{-6}$ is the ratio between the centrifugal and rotational constants of the molecule. As demonstrated in Fig.\ref{fig-dynamics}(\textbf{b}) for oxygen superrotors ($N=69$), even in light centrifuged molecules with strong chemical bond, the revival period may deviate by as much as 10\% from that predicted by the rigid-rotor model.
\begin{figure}[b]
\includegraphics[width=1\columnwidth]{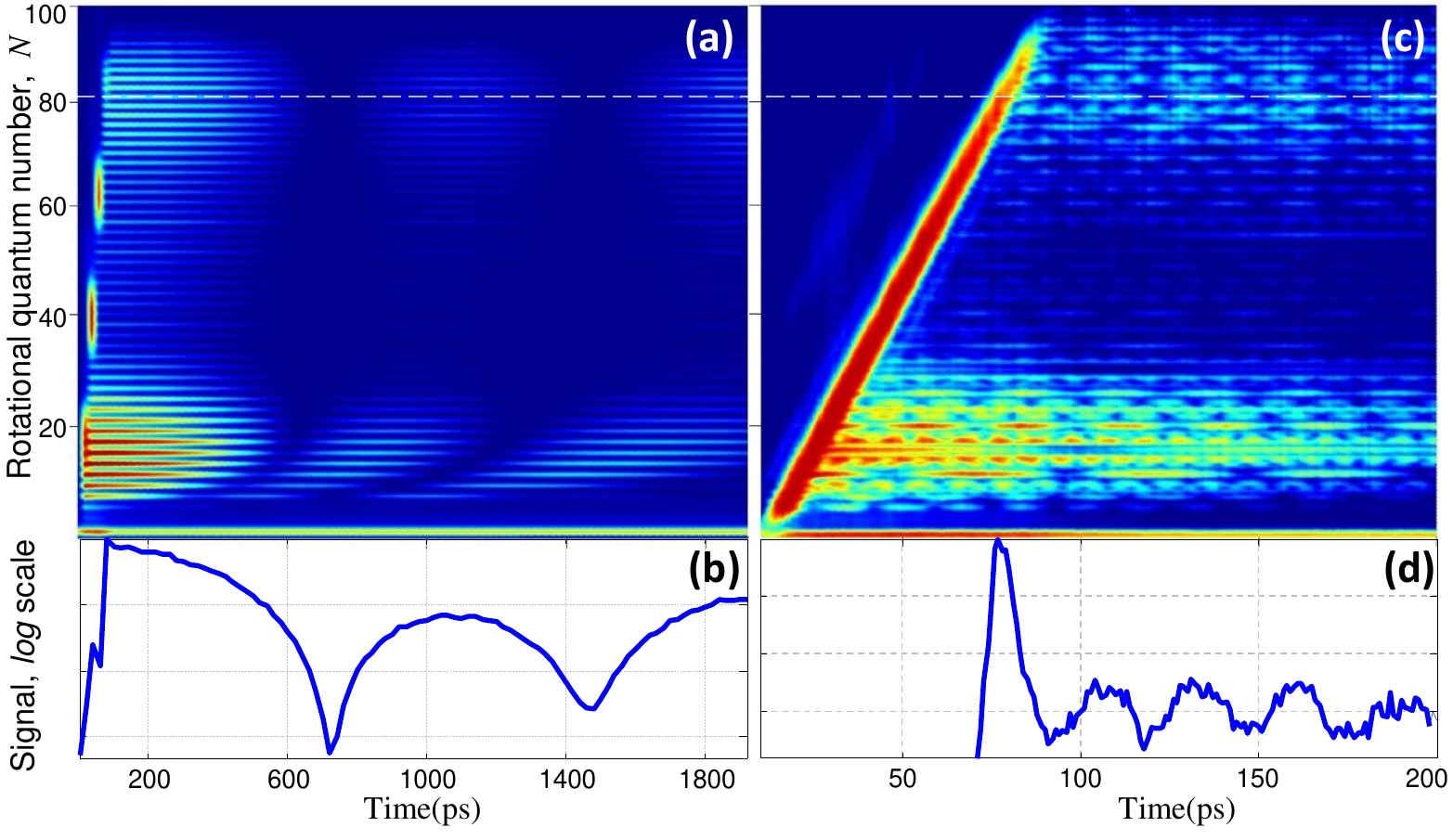}
\caption{(Color online) (\textbf{a,c}) Experimentally detected Raman spectrogram of centrifuged oxygen and centrifuged air, respectively, showing the rotational Raman spectrum as a function of the time delay between the beginning of the centrifuge pulse and the arrival of the probe pulse. Color coding is used to reflect the signal strength in logarithmic scale. (\textbf{b,d}) Cross-sections of the two-dimensional spectrogram at $N=81$ (horizontal dashed line), showing the spin-rotation oscillations and the oxygen-nitrogen beats in centrifuged air, respectively. Note different time scales for the left and right panels.}
\label{fig-interference}
\end{figure}

\subsubsection{Effects of spin-rotational coupling.}
Recent studies of collisional relaxation at extreme angular frequencies have shown that molecular superrotors are more immune to rotational decoherence than molecules in thermal ensembles \cite{Milner14a}. The effect stems from the increased adiabaticity of the collision process with increasing angular momentum. Higher robustness with respect to collisions and correspondingly longer time of rotational coherence offer new opportunities for high-resolution spectroscopy in dense gases under ambient conditions. In Fig.\ref{fig-interference}(\textbf{a}), fifty individually resolved rotational Raman lines of centrifuged O$_{2}$ are followed in time for as long as two nanoseconds \cite{Milner14b}. The lines reveal slow oscillations which originate from a weak coupling between the spin of the two unpaired electrons and the magnetic field of the rotating nuclei. Because of this spin-rotation coupling, any $N \rightarrow (N+2)$ Raman transition in oxygen consists of six separate lines belonging to one $Q$, two $R$ and three $S$ branches with $\Delta J=0,1$ and 2, respectively, where $J$ is the total rotational quantum number. The strength of both $Q$ and $R$ branches drops quickly with increasing $N$ and becomes negligibly small at $N>5$ \cite{Berard83}. The three stronger $S$ branches are split by less than 0.05 cm$^{-1}$, which corresponds to the oscillation period of about 1000 ps, clearly visible in the spectrogram (an example for $N=81$ is shown in Fig.\ref{fig-interference}(\textbf{b})).

\subsubsection{Rotational dynamics in gas mixtures.}
A Raman spectrogram for centrifuged ambient air is shown in Fig.\ref{fig-interference}(\textbf{c}). In contrast to the oscillatory signal from pure oxygen, multiple beat notes in the response from a mixture of O$_{2}$ and N$_{2}$ are due to the optical interference between coherently scattered light from the two simultaneously excited molecular species. For instance, the frequency of the $N=79 \rightarrow N=81$ Raman transition in oxygen is about 0.5 cm$^{-1}$ away from the one between $N=55$ and $N=57$ in nitrogen. If the spectral bandwidth of probe pulses exceeds this frequency difference, as is the case here, the two Raman signals interfere, producing the oscillations shown in Fig.\ref{fig-interference}(\textbf{d}).

\section{Summary and future directions}
In the recent work reviewed in this chapter, we have shown the creation and properties of molecular superrotors, illustrated by application of the optical centrifuge to simple diatomic molecules. The technique is certainly not limited to diatomics, and to date we have created superrotor wave packets in a range of molecules. All that is required is an anisotropic polarizability such that the molecule can be trapped in the rotating field or aligned by the pulsed field, and that is a property of virtually all molecules. The final rotational energy of the molecule is independent of molecular properties in most cases, and depends only on the bandwidth of the laser pulses. This means virtually any molecule can be put into a wave packet of extreme rotational states. This ability to excite molecules to extreme rotational states opens many possibilities for future investigations. For scattering studies, both molecule-molecule and molecule-surface, the role of both rotation and planar confinement can now be explored. Particularly interesting is the question of whether rotational energy can be transferred in collisions with a surface, and how the accommodation of rotational energy will depend on the rotational orientation with respect to the surface. The role of steric effects in molecular collisions and reaction dynamics leads to the possibility of controlling reactions through orientation effects or rotational hindrance of reaction channels.

Also appealing are the studies of unimolecular processes, because of the unusual properties of molecular superrotors. It has already been demonstrated that it is possible to break the chemical bonds through centrifugal forces \cite{Villeneuve00}, so the possibility exists that bonds can be selectively broken in a molecule based on differential centrifugal forces within a polyatomic superrotor. Beyond bond breaking, there is also the possibility of controlling the bending-driven isomerization in a molecule based on the centrifugal distortion created by high rotational energy. Conical intersections and other symmetry-related excited state interactions are very important in polyatomic photodissociation dynamics, so the question of how the significant distortions induced by extreme rotation will affect these dynamics could prove to be very interesting. In addition to dynamics, there will be interesting spectroscopic properties of molecules in extreme rotational states, such as rotation induced magnetism and emission of THz radiation by rotating polar molecules. Particularly interesting will be the spectroscopy of superrotors in superfluid helium nanodroplets, where it will be possible to study superfluid dynamics by measuring the high resolution REMPI spectrum of the embedded molecules as a function of rotational energy.

The above mentioned examples demonstrate that the ability to control and detect extreme molecular rotation opens up a very fruitful area of investigation in molecular dynamics and spectroscopy.


\end{document}